\documentclass[aip,reprint]{revtex4-1}
\draft
\usepackage{graphicx}
\usepackage{amsmath}
\begin{document}

\title{Simultaneous Matching of Phase and Amplitude for Spontaneous Parametric Down-conversion in Semiconductor Waveguides} 



\author{Albert Peralta Amores \& Marcin Swillo}
\email{marcin@kth.se}
\affiliation{School of Engineering Sciences, Royal Institute of Technology (KTH),  114-21 Stockholm, Sweden
}
\date{\today}
\begin{abstract}
We propose a non-uniform modulation of $\chi^{(2)}_{xyz}$ to significantly enhance photon pair generation efficiency via spontaneous parametric down-conversion in modal phase-matched semiconductor waveguides. This approach enables amplitude-matching in the transverse direction while preserving the phase-matching along the waveguide propagation axis. Our analysis predicts a tenfold efficiency increase in comparison to the most efficient non-modulated waveguide, and up to 13 orders of magnitude efficiency enhancements relative to solely phase-matched waveguides. Furthermore, we explore the implementation of a highly efficient compact twin-photon source, tunable across the communication band, using an amplitude- and phase-matched structure.

\end{abstract}
\maketitle 
Since the verification of Bell's tests using spontaneous-parametric down-conversion (SPDC) photon pair sources in 1988\cite{PhysRevLett.61.2921,PhysRevLett.61.50}, SPDC has emerged as a cornerstone in the generation of entangled photon pairs and squeezed states of light\cite{PhysRevD.23.1693, clauser1978}. Its pivotal role extends across multiple quantum communication schemes, including quantum key distribution (QKD) \cite{PhysRevLett.67.661}, teleportation \cite{Marcikic2003LongdistanceTO} and quantum repeaters \cite{PhysRevLett.81.5932}, all of which rely fundamentally on entanglement. Squeezed light finds applications in high precision sensing \cite{LIGO}, QKD \cite{Gehring_2015} and photonic quantum computation \cite{10.1063/1.5100160}.

Several solutions for phase-matching (PM) the SPDC process, making it efficient, have been proposed. The compensation of the chromatic dispersion by exploiting the birefringent properties of a nonlinear crystal, known as birefringence PM, was firstly demonstrated for SPDC in 1995\cite{PhysRevLett.75.4337}. The spatial modulation of the non-linear properties of a crystal along the propagation direction by reversing the non-linear interaction after a coherence length to achieve larger conversion efficiency, known as as quasi-PM, was introduced in 1962 \cite{PhysRev.127.1918}. Multiple methods such as temperature fluctuations during or after crystal growth\cite{10.1063/1.92035, 10.1063/1.107317}, electric fields \cite{10.1063/1.1657832} or direct wafer bonding of inverted domains\cite{10.1063/1.113370} have been implemented.

When the nonlinear media is structured in a waveguide (WG), equiphase velocity propagation between different guided modes, known as modal-PM, can be achieved through dispersion engineering \cite{10.1063/1.353040}. This has led to efficiency enhancements of several orders of magnitude with simultaneous bandwidth narrowing due to the strong spatial confinement of guided modes\cite{Fiorentino:07} while enabling the use of non-birefringent materials. In particular, III-V semiconductors exhibit larger refractive indexes \cite{PhysRevB.27.985} and $\chi^{(2)}_{xyz}$ susceptibilities \cite{Shoji:97} when compared to other traditionally used materials such as KTP or LiNbO$_3$ that exhibit birefringence.

\begin{figure}[h!]
\centering
\includegraphics[width=.43 \textwidth]{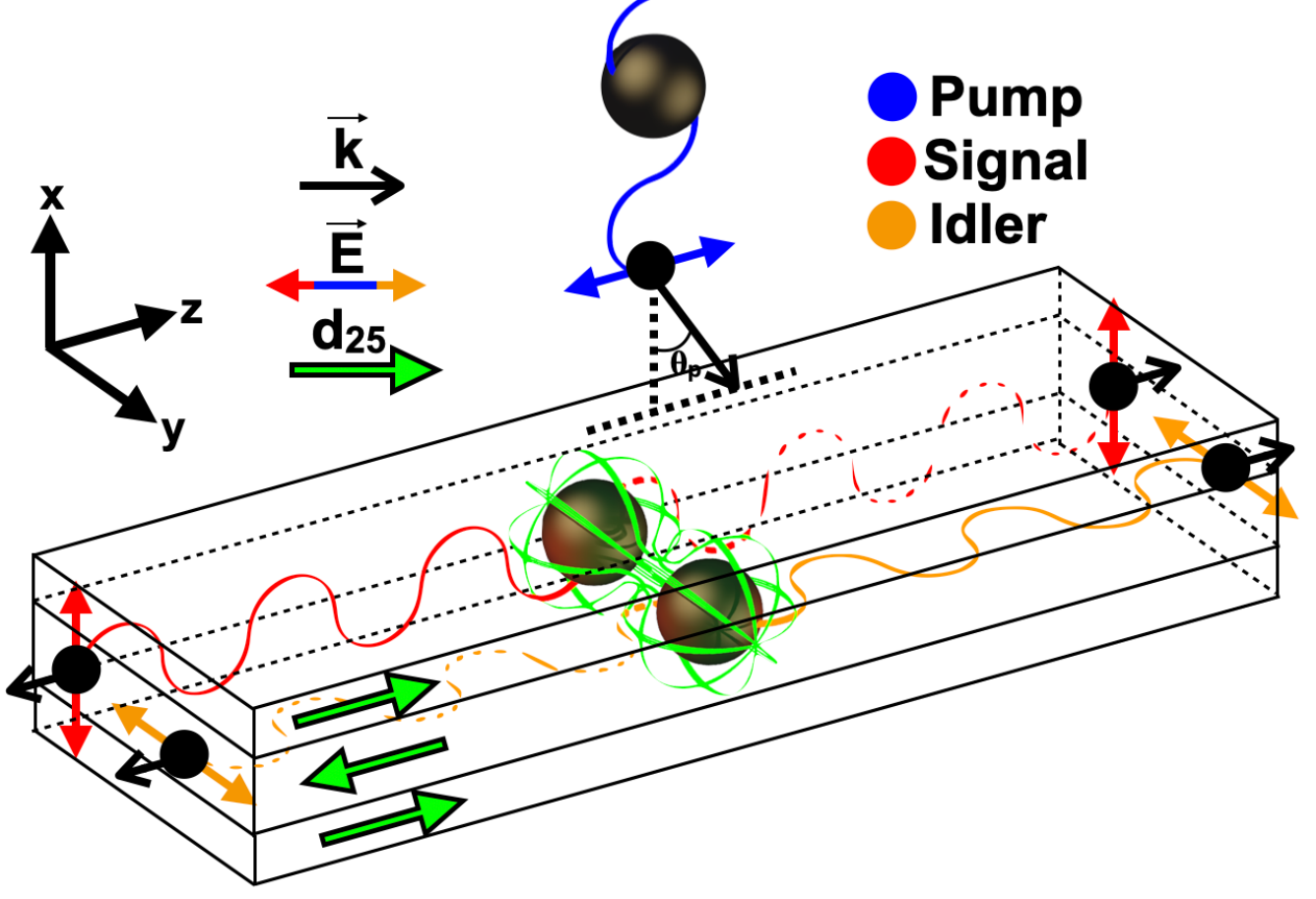}
\caption{Schematic of the APMS for the generation of counterpropagating photon pairs.}
\label{figconf}
\end{figure}

Here, we propose to significantly enhance the efficiency of SPDC processes by introducing a non-homogeneous spatial modulation of $\chi^{(2)}_{xyz}$ along the transverse plane of a modal-PM WG. As a result, efficiency enhancements up to 10$^{13}$ compared to identical non-modulated structures that retain the PM properties have been found. The most efficient PM WG exhibits a tenfold SPDC efficiency enhancement by introducing the modulation, shown in Fig.\ref{figconf}, resulting in a simultaneously amplitude- and PM structure (APMS). The presented approach enables further efficiency enhancements of PM structures (PMSs), previously considered optimized, making APMSs of considerable interest for the future development of integrated quantum photonic devices.

Consider a crystalline WG exhibiting $\overline{4}3m$ symmetry with propagation direction along its crystallographic axis ($\hat{z}$), as depicted in Fig.1. When a pump wave with polarization parallel to the crystallographic axis is focused along the WG, the $\chi^{(2)}_{xyz}$ tensor describes the spontaneous splitting of a pump photon into two photons with orthogonal polarization, as defined by the second-order nonlinear-polarization density function:
\begin{equation}
P^{(2)}_z(\textbf{r},t)=2d_{25}\varepsilon_0 E_y(\textbf{r},t)E_x(\textbf{r},t),
\end{equation}
where $d_{25}$ is the nonlinear coefficient of the material ($\mathbf{d}=\epsilon_0\chi^{(2)}_{xyz}$) and $E_y$, $E_x$ are the electric fields of the generated photons, respectively.

The SPDC process efficiency would then be governed by the PM condition $\mathbf{k}_p-\mathbf{k}_s-\mathbf{k}_i=0$, where the subscripts p, s and i account for the wave vectors of the pump, signal and idler, respectively. The generated signal and idler will excite TE- and TM-like guided modes within the WG, which, together with the lack of phase mismatch in the direction perpendicular to the WG, reduces the PM condition within the WG plane\cite{doi:10.1021/acsanm.1c04202}, yielding:
\begin{equation}
{k}_p \cdot sin(\theta_{p})={k}_{TM}+{k}_{TE},
\end{equation}
where $\theta_{inc}$ is the pump incidence angle. Pump tilting allows to compensate the momentum mismatch exhibited by the excited modes due to the WG form factor, enabling degenerated counterpropagating photon pairs generation. When assuming a pump coherence time longer than the propagation time of idler and signal photons ($\approx$ 12.5 ps/mm), the photon-pair generation rate in a WG of length L can be computed as:
\begin{equation}
\frac{\langle N \rangle}{t}= I_p\frac{8 \pi^2 L v_{TE,s} n_{TM,i}}{c_0 \lambda_s \lambda_i \epsilon_0  n_{c,i}^2 n_{TE,s}} |\eta|^2,
\label{rate}
\end{equation}
where I$_{p}$ is the pump power density and the parameters: $v_{TE}$, $n_{TE}$, $n_{TM}$, $c_0$, $\epsilon_0$ and $n_c$ are the group velocity and effective refractive index of the excited modes, speed of light in vacuum, vacuum permittivity and refractive index of the WG core, respectively.
The interacting fields overlap, $|\eta|^2$, greatly influences the photon pair generation rate and can be computed as:
\begin{equation}
\eta=\iint_W d(x) E_{TE}(x,y) E_{TM}(x,y) E_P (x,y) \,dx\,dy.
\label{overlap}
\end{equation}
Being; $d$, the InGaP nonlinear coefficienct (114 pm/V), $E_{TE}$ and $E_{TM}$, the normalized profiles of the transverse electric fields and $E_P(x,y)$ the pump beam amplitude profile in the WG with cross-section W, respectively. 

\begin{figure}[h!]
\centering
\includegraphics[width=.45 \textwidth]{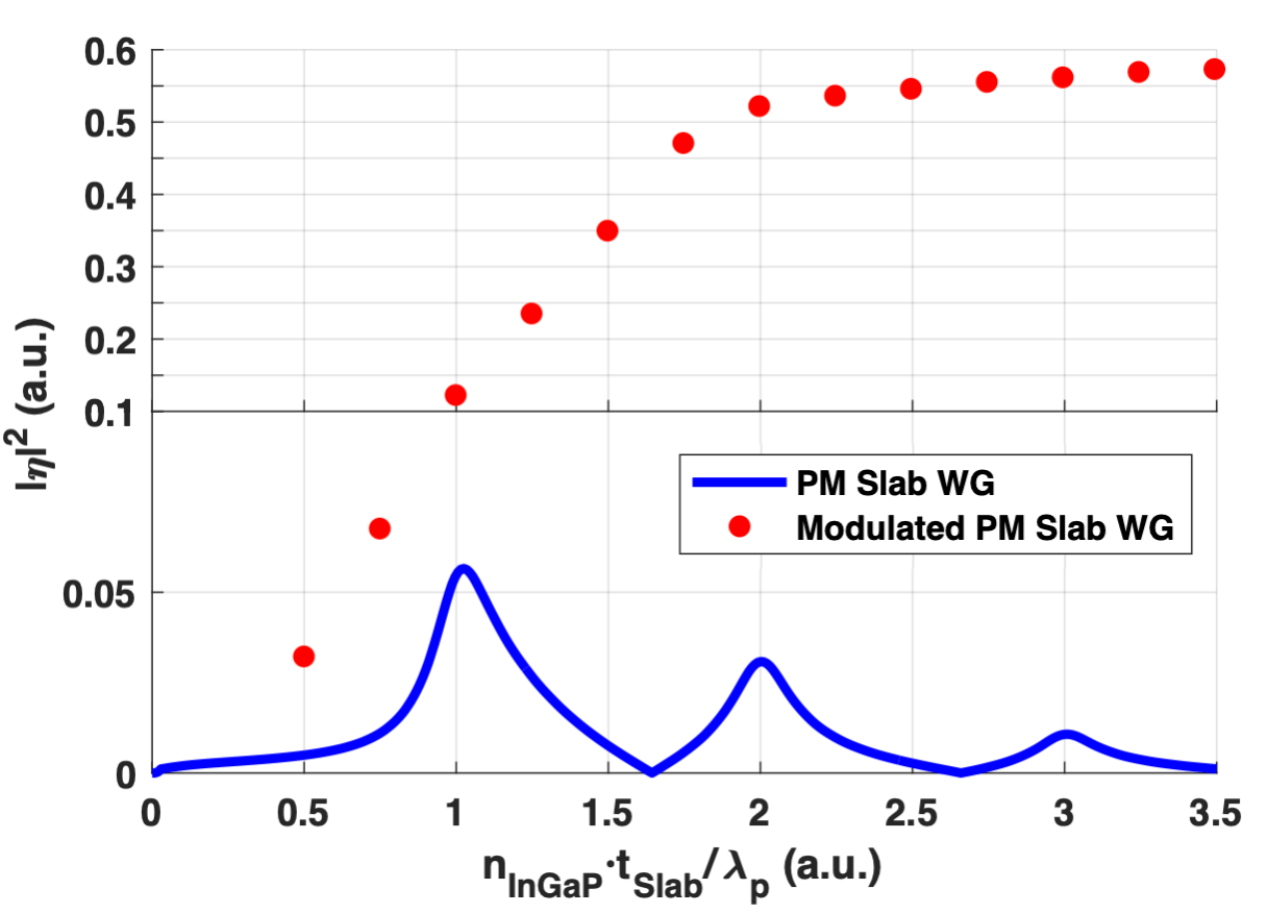}
                                                                            \caption{Interacting fields overlap,$|\eta|^2$, of a PM varying thickness InGaP slab WG with $\lambda_{p}$=726 nm before and after introducing the proposed modulation. (d(x) has been set to 1)}
\label{figslab}
\end{figure}

As shown in Fig.\ref{figslab} for an InGaP slab WG, $|\eta|^2$ exhibits local maxima when the WG thickness is a multiple of the pump wavelength, being the global maximum at one pump wavelength thickness. The maximum obtainable $|\eta|^2$ is $\approx$ 0.06 $|d|^2$, and threefold efficiency enhancements are achievable for the most efficient PMS by implementing the equivalent APMS.

Consider a 1 mm long 3 $\mu$m x 3 $\mu$m square WG with $\overline{4}3m$ symmetry and a refractive index, n, of 3.5 and 3.2 at $\lambda_{p}$=726 nm and $\lambda_{SPDC}$=1526 nm, respectively. Square WGs do not exhibit form birefringence, ensuring that the resulting PM condition (eq.2) is readily satisfied at degenerated wavelengths. The pump amplitude profile, E$_p$, will determine the efficiency of the SPDC process.
When $\lambda_{PUMP}$ = 726 nm, the WG accommodates 29 half pump oscillations. The product of the interacting field profiles along the WG vertical cross-section, shown in Fig.3, results in an interacting fields overlap $|\eta|^2=9\cdot10^{-15}\cdot|d|^2$ due to the $\pi$ phase shift at subsequent antinodes and the pump electric field profiles interfering destructively at the center of the WG. Inverting the sign of the nonlinear coefficient, $d_{25}$, to compensate the $\pi$ phase shift results in the additive contribution of the entire nonlinear media to the SPDC process ($|\eta|^2=0.17\cdot|d|^2$). Consequently, the PMS and the APMS exhibit an efficiency of 2.74$\cdot10^{-15}$ and 0.052 photon pairs per pump photon, respectively.
\begin{figure}[h!]
\centering
\includegraphics[width=.4\textwidth]{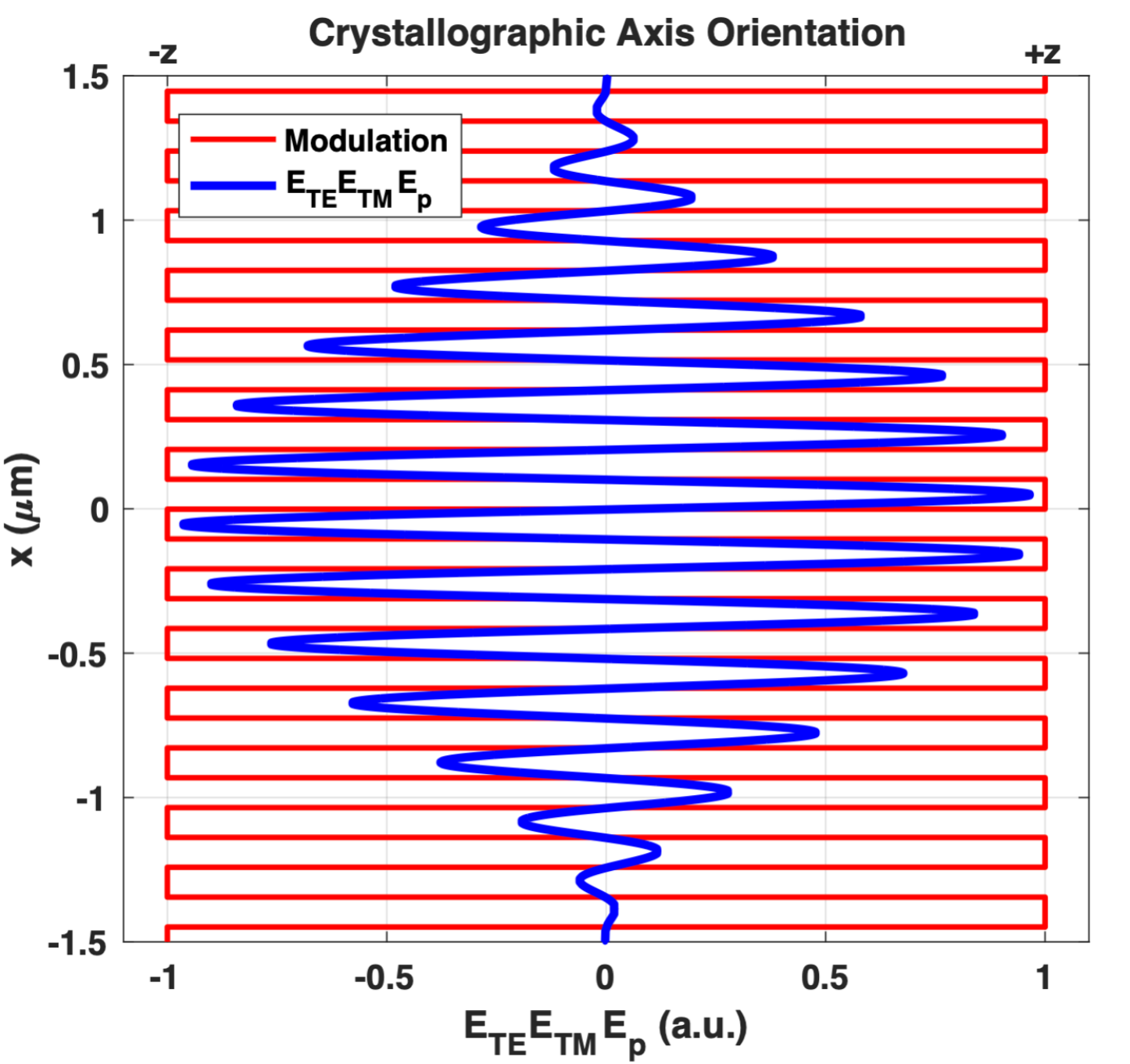}
\caption{Product of the the pump profile and the transverse electric field profiles of the guided TE and TM modes and the proposed non-periodic modulation of $\chi^{(2)}_{xyz}$ to achieve a constructive contribution of the entire nonlinear volume to the SPDC process.}
\label{figcorr}
\end{figure} 

\begin{figure}[h!]
\centering
\includegraphics[width=.48 \textwidth]{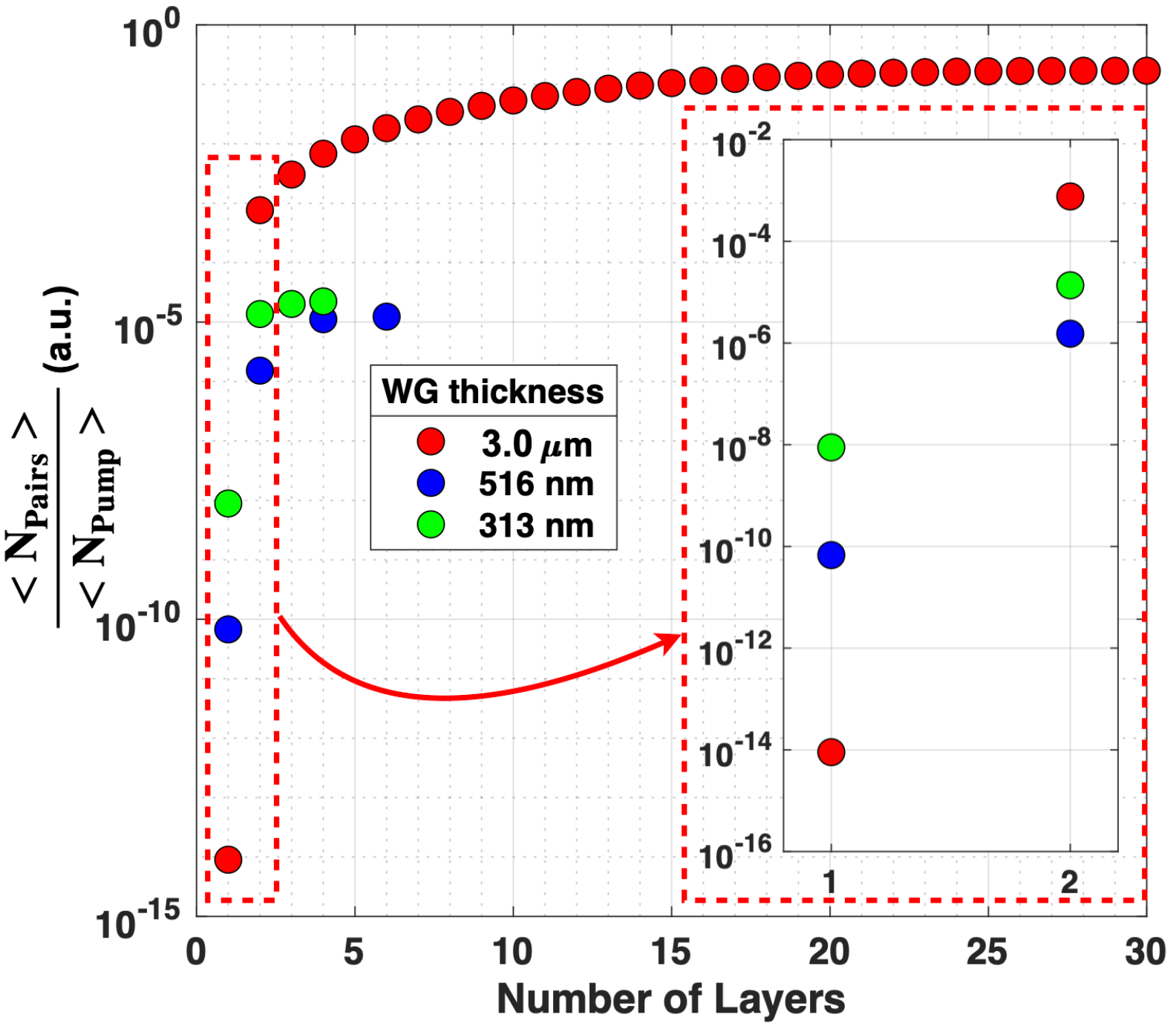}
\caption{Photon-pair generation efficiency as a function of the number of modulation layers introduced, being one layer the PMS and the maximum number of layers the APMS.}
\label{figwidth}
\end{figure}

The proposed modulation is particularly interesting for WGs which exhibit a thickness close to an odd number of half $\lambda_{p}$, inherently less efficient due to the pump destructive interference at the center of the WG, resulting in an out-of-phase contribution from both WG halves. In such cases, the implementation of a quasi-APMS consisting of two layers with inverted crystallographic axis results in a substantial efficiency enhancement while notably easifying its implementation. A lower bound of $10^3$ efficiency enhancement has been found for a WG with a thickness of 3 half $\lambda_{p}$ while a $10^{10}$ enhancement has been found for a WG of 29 half $\lambda_{p}$ thickness, as shown in Fig.\ref{figwidth}. Furthermore, quasi-APMS enables the implementation of $\mu$m size geometries with reduced number of required layers, easing their implementation while enjoying the largest efficiency enhancement, even surpassing compact geometries which are inherently more efficient when solely PM.

\begin{figure}[h!]
\centering
\includegraphics[width=.48 \textwidth]{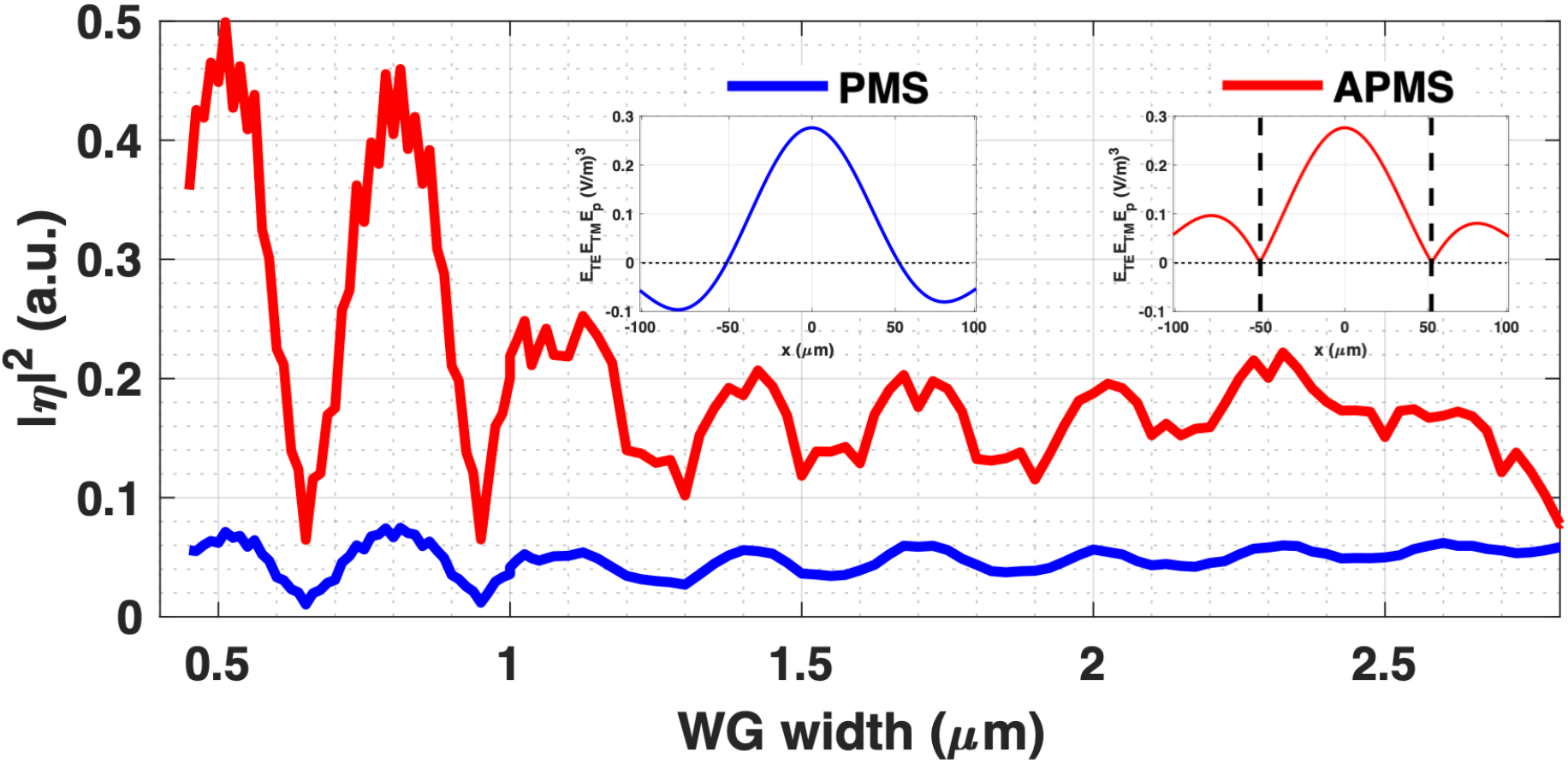}
\caption{Interacting fields overlap of a 207 nm thick InGaP waveguide for the PMS and the APMS as a function of the WG width.}
\label{figlast}
\end{figure}

The WG width determines the pump amplitude profile along y and the excited guided modes within the WG. The SPDC process efficiency as a function of the WG width is shown in Fig.\ref{figlast} for the most efficient PMS, the WG which accommodates a single pump wavelength (207 nm thickness). As a result, a central layer with half a pump wavelength optical thickness (103.5 nm) has to be inverted in order to achieve an additive contribution of the entire nonlinear media, as depicted in Fig.5.(Inset). Photon pair generation efficiency improvements up to $\approx 9.5$ have been found when comparing the APMS and the PMS. 

\begin{figure}[h!]
\centering
\includegraphics[width=0.5 \textwidth]{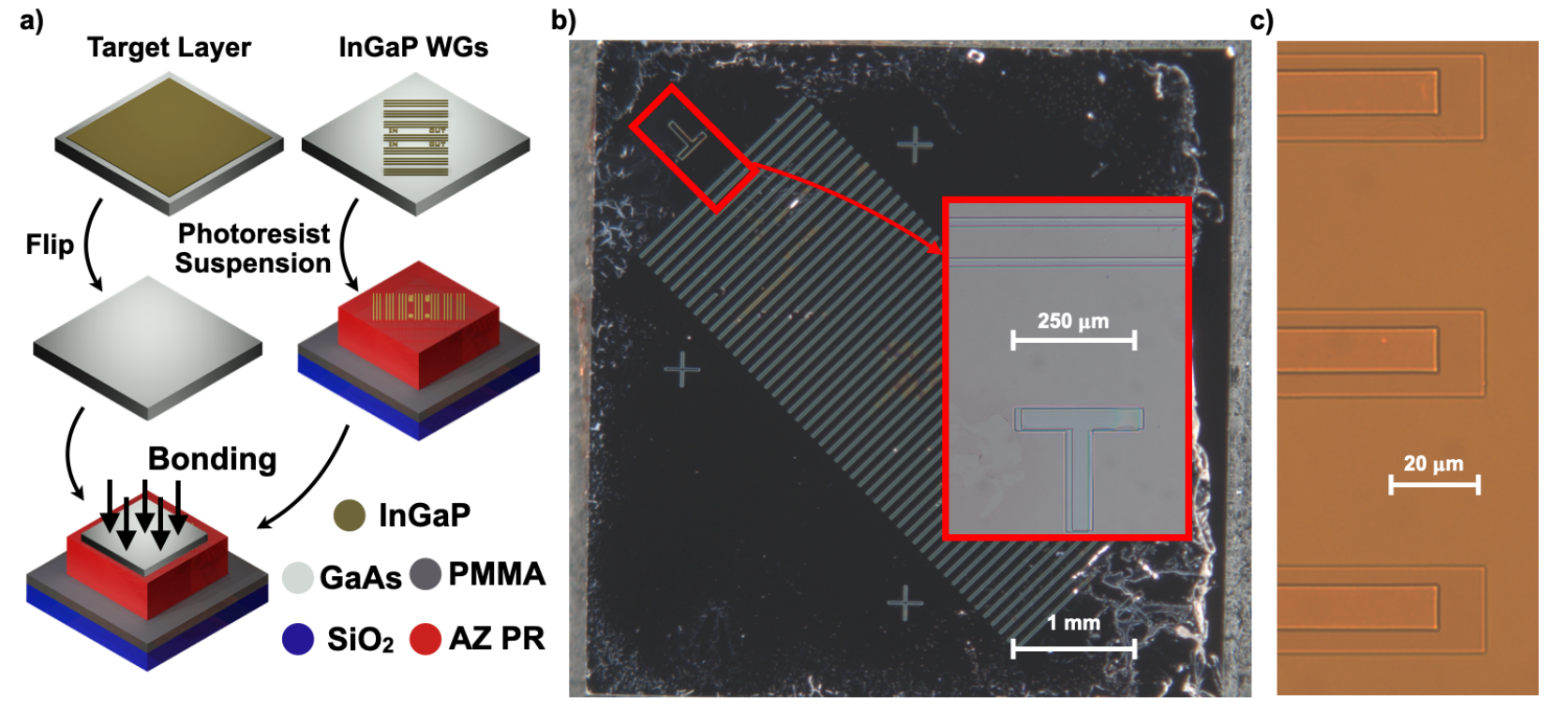}
\caption{Outline of the proposed implementation process (a). A 100 nm thick InGaP WG array transfered to a 5 x 5 mm InGaP sample after rotating the crystallographic axis 90º (b). Two 100 nm InGaP layers with 20 $\mu$m and 15 $\mu$m width on top of an InGaP layer (c).}
\label{INGAPBOND}
\end{figure}

The APMS is fully compatible with a recently demonstrated counterpropagating twin photon source that exhibits tunable degenerated emission from 1400 to 1600 nm with $\approx10^{-5}$ efficiency\cite{PhysRevA.110.063713}. The transfer of 250 nm thick InGaP WGs on fussed silica is extensively described in\cite{Amores:24}. Here, we demonstrate the implementation an InGaP layer with a central region with inverted crystallographic axis. Through minimal modifications, outlined in Fig.\ref{INGAPBOND}a, the successful transfer of an array of 100 nm thick and 2 mm long InGaP WGs on top of 5 x 5 mm InGaP layer after rotating the crystallographic axis 90° via native oxide molecular bonding is shown in Fig.\ref{INGAPBOND}b. The subsequent transfer, resulting in two alternated layers of 100 nm on top of InGaP is shown in Fig.\ref{INGAPBOND}c.

To conclude, a spatial modulation of $\chi^{(2)}_{xyz}$ to further enhance the efficiency of PMSs exhibiting -43m symmetry has proposed. The presented approach enables the simultaneous amplitude- and PM, resulting in an efficient utilization of the entire nonlinear volume. As a result, photon pair generation via SPDC efficiency enhancements up to $10^{13}$ have been found when comparing APMSs and PMSs. The compatibility with recently demonstrated counterpropagating twin photon sources and the feasibility of the InGaP self-bonding through native oxide after rotating the crystallographic axis 90º, makes APMSs a promising approach for the future developement of integrated non-classical sources.
\begin{acknowledgments}
We acknowledge the financial support from the Knut and Alice Wallenberg Foundation through the Wallenberg Center for Quantum Technology (WACQT). Financial support from the Swedish Research Council VR (Grant No. 2018-03457) is also acknowledged.
\end{acknowledgments}
\bibliography{references}

\begin{thebibliography}{23}%
\makeatletter
\providecommand \@ifxundefined [1]{%
 \@ifx{#1\undefined}
}%
\providecommand \@ifnum [1]{%
 \ifnum #1\expandafter \@firstoftwo
 \else \expandafter \@secondoftwo
 \fi
}%
\providecommand \@ifx [1]{%
 \ifx #1\expandafter \@firstoftwo
 \else \expandafter \@secondoftwo
 \fi
}%
\providecommand \natexlab [1]{#1}%
\providecommand \enquote  [1]{``#1''}%
\providecommand \bibnamefont  [1]{#1}%
\providecommand \bibfnamefont [1]{#1}%
\providecommand \citenamefont [1]{#1}%
\providecommand \href@noop [0]{\@secondoftwo}%
\providecommand \href [0]{\begingroup \@sanitize@url \@href}%
\providecommand \@href[1]{\@@startlink{#1}\@@href}%
\providecommand \@@href[1]{\endgroup#1\@@endlink}%
\providecommand \@sanitize@url [0]{\catcode `\\12\catcode `\$12\catcode `\&12\catcode `\#12\catcode `\^12\catcode `\_12\catcode `\%12\relax}%
\providecommand \@@startlink[1]{}%
\providecommand \@@endlink[0]{}%
\providecommand \url  [0]{\begingroup\@sanitize@url \@url }%
\providecommand \@url [1]{\endgroup\@href {#1}{\urlprefix }}%
\providecommand \urlprefix  [0]{URL }%
\providecommand \Eprint [0]{\href }%
\providecommand \doibase [0]{http://dx.doi.org/}%
\providecommand \selectlanguage [0]{\@gobble}%
\providecommand \bibinfo  [0]{\@secondoftwo}%
\providecommand \bibfield  [0]{\@secondoftwo}%
\providecommand \translation [1]{[#1]}%
\providecommand \BibitemOpen [0]{}%
\providecommand \bibitemStop [0]{}%
\providecommand \bibitemNoStop [0]{.\EOS\space}%
\providecommand \EOS [0]{\spacefactor3000\relax}%
\providecommand \BibitemShut  [1]{\csname bibitem#1\endcsname}%
\let\auto@bib@innerbib\@empty
\bibitem [{\citenamefont {Shih}\ and\ \citenamefont {Alley}(1988)}]{PhysRevLett.61.2921}%
  \BibitemOpen
  \bibfield  {author} {\bibinfo {author} {\bibfnamefont {Y.~H.}\ \bibnamefont {Shih}}\ and\ \bibinfo {author} {\bibfnamefont {C.~O.}\ \bibnamefont {Alley}},\ }\bibfield  {title} {\enquote {\bibinfo {title} {New type of einstein-podolsky-rosen-bohm experiment using pairs of light quanta produced by optical parametric down conversion},}\ }\href {\doibase 10.1103/PhysRevLett.61.2921} {\bibfield  {journal} {\bibinfo  {journal} {Phys. Rev. Lett.}\ }\textbf {\bibinfo {volume} {61}},\ \bibinfo {pages} {2921--2924} (\bibinfo {year} {1988})}\BibitemShut {NoStop}%
\bibitem [{\citenamefont {Ou}\ and\ \citenamefont {Mandel}(1988)}]{PhysRevLett.61.50}%
  \BibitemOpen
  \bibfield  {author} {\bibinfo {author} {\bibfnamefont {Z.~Y.}\ \bibnamefont {Ou}}\ and\ \bibinfo {author} {\bibfnamefont {L.}~\bibnamefont {Mandel}},\ }\bibfield  {title} {\enquote {\bibinfo {title} {Violation of bell's inequality and classical probability in a two-photon correlation experiment},}\ }\href {\doibase 10.1103/PhysRevLett.61.50} {\bibfield  {journal} {\bibinfo  {journal} {Phys. Rev. Lett.}\ }\textbf {\bibinfo {volume} {61}},\ \bibinfo {pages} {50--53} (\bibinfo {year} {1988})}\BibitemShut {NoStop}%
\bibitem [{\citenamefont {Caves}(1981)}]{PhysRevD.23.1693}%
  \BibitemOpen
  \bibfield  {author} {\bibinfo {author} {\bibfnamefont {C.~M.}\ \bibnamefont {Caves}},\ }\bibfield  {title} {\enquote {\bibinfo {title} {Quantum-mechanical noise in an interferometer},}\ }\href {\doibase 10.1103/PhysRevD.23.1693} {\bibfield  {journal} {\bibinfo  {journal} {Phys. Rev. D}\ }\textbf {\bibinfo {volume} {23}},\ \bibinfo {pages} {1693--1708} (\bibinfo {year} {1981})}\BibitemShut {NoStop}%
\bibitem [{\citenamefont {Clauser}(1978)}]{clauser1978}%
  \BibitemOpen
  \bibfield  {author} {\bibinfo {author} {\bibfnamefont {J.~F.}\ \bibnamefont {Clauser}},\ }\bibfield  {title} {\enquote {\bibinfo {title} {Bell's theorem. experimental tests and implications},}\ }\href {\doibase 10.1088/0034-4885/41/12/002} {\bibfield  {journal} {\bibinfo  {journal} {Reports on Progress in Physics}\ }\textbf {\bibinfo {volume} {41}},\ \bibinfo {pages} {1881} (\bibinfo {year} {1978})}\BibitemShut {NoStop}%
\bibitem [{\citenamefont {Ekert}(1991)}]{PhysRevLett.67.661}%
  \BibitemOpen
  \bibfield  {author} {\bibinfo {author} {\bibfnamefont {A.~K.}\ \bibnamefont {Ekert}},\ }\bibfield  {title} {\enquote {\bibinfo {title} {Quantum cryptography based on bell's theorem},}\ }\href {\doibase 10.1103/PhysRevLett.67.661} {\bibfield  {journal} {\bibinfo  {journal} {Phys. Rev. Lett.}\ }\textbf {\bibinfo {volume} {67}},\ \bibinfo {pages} {661--663} (\bibinfo {year} {1991})}\BibitemShut {NoStop}%
\bibitem [{\citenamefont {Marcikic}\ \emph {et~al.}(2003)\citenamefont {Marcikic}, \citenamefont {de~Riedmatten}, \citenamefont {Tittel}, \citenamefont {Tittel}, \citenamefont {Zbinden},\ and\ \citenamefont {Gisin}}]{Marcikic2003LongdistanceTO}%
  \BibitemOpen
  \bibfield  {author} {\bibinfo {author} {\bibfnamefont {I.}~\bibnamefont {Marcikic}}, \bibinfo {author} {\bibfnamefont {H.}~\bibnamefont {de~Riedmatten}}, \bibinfo {author} {\bibfnamefont {W.}~\bibnamefont {Tittel}}, \bibinfo {author} {\bibfnamefont {W.}~\bibnamefont {Tittel}}, \bibinfo {author} {\bibfnamefont {H.}~\bibnamefont {Zbinden}}, \ and\ \bibinfo {author} {\bibfnamefont {N.}~\bibnamefont {Gisin}},\ }\bibfield  {title} {\enquote {\bibinfo {title} {Long-distance teleportation of qubits at telecommunication wavelengths},}\ }\href {https://api.semanticscholar.org/CorpusID:4303767} {\bibfield  {journal} {\bibinfo  {journal} {Nature}\ }\textbf {\bibinfo {volume} {421}},\ \bibinfo {pages} {509--513} (\bibinfo {year} {2003})}\BibitemShut {NoStop}%
\bibitem [{\citenamefont {Briegel}\ \emph {et~al.}(1998)\citenamefont {Briegel}, \citenamefont {D\"ur}, \citenamefont {Cirac},\ and\ \citenamefont {Zoller}}]{PhysRevLett.81.5932}%
  \BibitemOpen
  \bibfield  {author} {\bibinfo {author} {\bibfnamefont {H.-J.}\ \bibnamefont {Briegel}}, \bibinfo {author} {\bibfnamefont {W.}~\bibnamefont {D\"ur}}, \bibinfo {author} {\bibfnamefont {J.~I.}\ \bibnamefont {Cirac}}, \ and\ \bibinfo {author} {\bibfnamefont {P.}~\bibnamefont {Zoller}},\ }\bibfield  {title} {\enquote {\bibinfo {title} {Quantum repeaters: The role of imperfect local operations in quantum communication},}\ }\href {\doibase 10.1103/PhysRevLett.81.5932} {\bibfield  {journal} {\bibinfo  {journal} {Phys. Rev. Lett.}\ }\textbf {\bibinfo {volume} {81}},\ \bibinfo {pages} {5932--5935} (\bibinfo {year} {1998})}\BibitemShut {NoStop}%
\bibitem [{\citenamefont {Collaboration}(2011)}]{LIGO}%
  \BibitemOpen
  \bibfield  {author} {\bibinfo {author} {\bibfnamefont {T.~L.~S.}\ \bibnamefont {Collaboration}},\ }\bibfield  {title} {\enquote {\bibinfo {title} {A gravitational wave observatory operating beyond the quantum shot-noise limit},}\ }\href {\doibase 10.1038/nphys2083} {\bibfield  {journal} {\bibinfo  {journal} {Nature Physics}\ }\textbf {\bibinfo {volume} {7}},\ \bibinfo {pages} {962--965} (\bibinfo {year} {2011})}\BibitemShut {NoStop}%
\bibitem [{\citenamefont {Gehring}\ \emph {et~al.}(2015)\citenamefont {Gehring}, \citenamefont {Händchen}, \citenamefont {Duhme}, \citenamefont {Furrer}, \citenamefont {Franz}, \citenamefont {Pacher}, \citenamefont {Werner},\ and\ \citenamefont {Schnabel}}]{Gehring_2015}%
  \BibitemOpen
  \bibfield  {author} {\bibinfo {author} {\bibfnamefont {T.}~\bibnamefont {Gehring}}, \bibinfo {author} {\bibfnamefont {V.}~\bibnamefont {Händchen}}, \bibinfo {author} {\bibfnamefont {J.}~\bibnamefont {Duhme}}, \bibinfo {author} {\bibfnamefont {F.}~\bibnamefont {Furrer}}, \bibinfo {author} {\bibfnamefont {T.}~\bibnamefont {Franz}}, \bibinfo {author} {\bibfnamefont {C.}~\bibnamefont {Pacher}}, \bibinfo {author} {\bibfnamefont {R.~F.}\ \bibnamefont {Werner}}, \ and\ \bibinfo {author} {\bibfnamefont {R.}~\bibnamefont {Schnabel}},\ }\bibfield  {title} {\enquote {\bibinfo {title} {Implementation of continuous-variable quantum key distribution with composable and one-sided-device-independent security against coherent attacks},}\ }\href {\doibase 10.1038/ncomms9795} {\bibfield  {journal} {\bibinfo  {journal} {Nature Communications}\ }\textbf {\bibinfo {volume} {6}} (\bibinfo {year} {2015}),\ 10.1038/ncomms9795}\BibitemShut {NoStop}%
\bibitem [{\citenamefont {Takeda}\ and\ \citenamefont {Furusawa}(2019)}]{10.1063/1.5100160}%
  \BibitemOpen
  \bibfield  {author} {\bibinfo {author} {\bibfnamefont {S.}~\bibnamefont {Takeda}}\ and\ \bibinfo {author} {\bibfnamefont {A.}~\bibnamefont {Furusawa}},\ }\bibfield  {title} {\enquote {\bibinfo {title} {{Toward large-scale fault-tolerant universal photonic quantum computing}},}\ }\href {\doibase 10.1063/1.5100160} {\bibfield  {journal} {\bibinfo  {journal} {APL Photonics}\ }\textbf {\bibinfo {volume} {4}},\ \bibinfo {pages} {060902} (\bibinfo {year} {2019})}\BibitemShut {NoStop}%
\bibitem [{\citenamefont {Kwiat}\ \emph {et~al.}(1995)\citenamefont {Kwiat}, \citenamefont {Mattle}, \citenamefont {Weinfurter}, \citenamefont {Zeilinger}, \citenamefont {Sergienko},\ and\ \citenamefont {Shih}}]{PhysRevLett.75.4337}%
  \BibitemOpen
  \bibfield  {author} {\bibinfo {author} {\bibfnamefont {P.~G.}\ \bibnamefont {Kwiat}}, \bibinfo {author} {\bibfnamefont {K.}~\bibnamefont {Mattle}}, \bibinfo {author} {\bibfnamefont {H.}~\bibnamefont {Weinfurter}}, \bibinfo {author} {\bibfnamefont {A.}~\bibnamefont {Zeilinger}}, \bibinfo {author} {\bibfnamefont {A.~V.}\ \bibnamefont {Sergienko}}, \ and\ \bibinfo {author} {\bibfnamefont {Y.}~\bibnamefont {Shih}},\ }\bibfield  {title} {\enquote {\bibinfo {title} {New high-intensity source of polarization-entangled photon pairs},}\ }\href {\doibase 10.1103/PhysRevLett.75.4337} {\bibfield  {journal} {\bibinfo  {journal} {Phys. Rev. Lett.}\ }\textbf {\bibinfo {volume} {75}},\ \bibinfo {pages} {4337--4341} (\bibinfo {year} {1995})}\BibitemShut {NoStop}%
\bibitem [{\citenamefont {Armstrong}\ \emph {et~al.}(1962)\citenamefont {Armstrong}, \citenamefont {Bloembergen}, \citenamefont {Ducuing},\ and\ \citenamefont {Pershan}}]{PhysRev.127.1918}%
  \BibitemOpen
  \bibfield  {author} {\bibinfo {author} {\bibfnamefont {J.~A.}\ \bibnamefont {Armstrong}}, \bibinfo {author} {\bibfnamefont {N.}~\bibnamefont {Bloembergen}}, \bibinfo {author} {\bibfnamefont {J.}~\bibnamefont {Ducuing}}, \ and\ \bibinfo {author} {\bibfnamefont {P.~S.}\ \bibnamefont {Pershan}},\ }\bibfield  {title} {\enquote {\bibinfo {title} {Interactions between light waves in a nonlinear dielectric},}\ }\href {\doibase 10.1103/PhysRev.127.1918} {\bibfield  {journal} {\bibinfo  {journal} {Phys. Rev.}\ }\textbf {\bibinfo {volume} {127}},\ \bibinfo {pages} {1918--1939} (\bibinfo {year} {1962})}\BibitemShut {NoStop}%
\bibitem [{\citenamefont {Feng}\ \emph {et~al.}(1980)\citenamefont {Feng}, \citenamefont {Ming}, \citenamefont {Hong}, \citenamefont {Yang}, \citenamefont {Zhu}, \citenamefont {Yang},\ and\ \citenamefont {Wang}}]{10.1063/1.92035}%
  \BibitemOpen
  \bibfield  {author} {\bibinfo {author} {\bibfnamefont {D.}~\bibnamefont {Feng}}, \bibinfo {author} {\bibfnamefont {N.}~\bibnamefont {Ming}}, \bibinfo {author} {\bibfnamefont {J.}~\bibnamefont {Hong}}, \bibinfo {author} {\bibfnamefont {Y.}~\bibnamefont {Yang}}, \bibinfo {author} {\bibfnamefont {J.}~\bibnamefont {Zhu}}, \bibinfo {author} {\bibfnamefont {Z.}~\bibnamefont {Yang}}, \ and\ \bibinfo {author} {\bibfnamefont {Y.}~\bibnamefont {Wang}},\ }\bibfield  {title} {\enquote {\bibinfo {title} {{Enhancement of second‐harmonic generation in LiNbO3 crystals with periodic laminar ferroelectric domains}},}\ }\href {\doibase 10.1063/1.92035} {\bibfield  {journal} {\bibinfo  {journal} {Applied Physics Letters}\ }\textbf {\bibinfo {volume} {37}},\ \bibinfo {pages} {607--609} (\bibinfo {year} {1980})},\ \Eprint {http://arxiv.org/abs/https://pubs.aip.org/aip/apl/article-pdf/37/7/607/18442220/607\_1\_online.pdf} {https://pubs.aip.org/aip/apl/article-pdf/37/7/607/18442220/607\_1\_online.pdf} \BibitemShut {NoStop}%
\bibitem [{\citenamefont {Mizuuchi}\ and\ \citenamefont {Yamamoto}(1992)}]{10.1063/1.107317}%
  \BibitemOpen
  \bibfield  {author} {\bibinfo {author} {\bibfnamefont {K.}~\bibnamefont {Mizuuchi}}\ and\ \bibinfo {author} {\bibfnamefont {K.}~\bibnamefont {Yamamoto}},\ }\bibfield  {title} {\enquote {\bibinfo {title} {{Highly efficient quasi‐phase‐matched second‐harmonic generation using a first‐order periodically domain‐inverted LiTaO3 waveguide}},}\ }\href {\doibase 10.1063/1.107317} {\bibfield  {journal} {\bibinfo  {journal} {Applied Physics Letters}\ }\textbf {\bibinfo {volume} {60}},\ \bibinfo {pages} {1283--1285} (\bibinfo {year} {1992})},\ \Eprint {http://arxiv.org/abs/https://pubs.aip.org/aip/apl/article-pdf/60/11/1283/18488051/1283\_1\_online.pdf} {https://pubs.aip.org/aip/apl/article-pdf/60/11/1283/18488051/1283\_1\_online.pdf} \BibitemShut {NoStop}%
\bibitem [{\citenamefont {Camlibel}(1969)}]{10.1063/1.1657832}%
  \BibitemOpen
  \bibfield  {author} {\bibinfo {author} {\bibfnamefont {I.}~\bibnamefont {Camlibel}},\ }\bibfield  {title} {\enquote {\bibinfo {title} {{Spontaneous Polarization Measurements in Several Ferroelectric Oxides Using a Pulsed‐Field Method}},}\ }\href {\doibase 10.1063/1.1657832} {\bibfield  {journal} {\bibinfo  {journal} {Journal of Applied Physics}\ }\textbf {\bibinfo {volume} {40}},\ \bibinfo {pages} {1690--1693} (\bibinfo {year} {1969})},\ \Eprint {http://arxiv.org/abs/https://pubs.aip.org/aip/jap/article-pdf/40/4/1690/18349886/1690\_1\_online.pdf} {https://pubs.aip.org/aip/jap/article-pdf/40/4/1690/18349886/1690\_1\_online.pdf} \BibitemShut {NoStop}%
\bibitem [{\citenamefont {Yoo}\ \emph {et~al.}(1995)\citenamefont {Yoo}, \citenamefont {Bhat}, \citenamefont {Caneau},\ and\ \citenamefont {Koza}}]{10.1063/1.113370}%
  \BibitemOpen
  \bibfield  {author} {\bibinfo {author} {\bibfnamefont {S.~J.~B.}\ \bibnamefont {Yoo}}, \bibinfo {author} {\bibfnamefont {R.}~\bibnamefont {Bhat}}, \bibinfo {author} {\bibfnamefont {C.}~\bibnamefont {Caneau}}, \ and\ \bibinfo {author} {\bibfnamefont {M.~A.}\ \bibnamefont {Koza}},\ }\bibfield  {title} {\enquote {\bibinfo {title} {{Quasi‐phase‐matched second‐harmonic generation in AlGaAs waveguides with periodic domain inversion achieved by wafer‐bonding}},}\ }\href {\doibase 10.1063/1.113370} {\bibfield  {journal} {\bibinfo  {journal} {Applied Physics Letters}\ }\textbf {\bibinfo {volume} {66}},\ \bibinfo {pages} {3410--3412} (\bibinfo {year} {1995})},\ \Eprint {http://arxiv.org/abs/https://pubs.aip.org/aip/apl/article-pdf/66/25/3410/18512110/3410\_1\_online.pdf} {https://pubs.aip.org/aip/apl/article-pdf/66/25/3410/18512110/3410\_1\_online.pdf} \BibitemShut {NoStop}%
\bibitem [{\citenamefont {Zhang}\ \emph {et~al.}(1993)\citenamefont {Zhang}, \citenamefont {Chandler}, \citenamefont {Townsend}, \citenamefont {Alwahabi}, \citenamefont {Pityana},\ and\ \citenamefont {McCaffery}}]{10.1063/1.353040}%
  \BibitemOpen
  \bibfield  {author} {\bibinfo {author} {\bibfnamefont {L.}~\bibnamefont {Zhang}}, \bibinfo {author} {\bibfnamefont {P.~J.}\ \bibnamefont {Chandler}}, \bibinfo {author} {\bibfnamefont {P.~D.}\ \bibnamefont {Townsend}}, \bibinfo {author} {\bibfnamefont {Z.~T.}\ \bibnamefont {Alwahabi}}, \bibinfo {author} {\bibfnamefont {S.~L.}\ \bibnamefont {Pityana}}, \ and\ \bibinfo {author} {\bibfnamefont {A.~J.}\ \bibnamefont {McCaffery}},\ }\bibfield  {title} {\enquote {\bibinfo {title} {{Frequency doubling in ion‐implanted KTiOPO4 planar waveguides with 25 conversion efficiency}},}\ }\href {\doibase 10.1063/1.353040} {\bibfield  {journal} {\bibinfo  {journal} {Journal of Applied Physics}\ }\textbf {\bibinfo {volume} {73}},\ \bibinfo {pages} {2695--2699} (\bibinfo {year} {1993})},\ \Eprint {http://arxiv.org/abs/https://pubs.aip.org/aip/jap/article-pdf/73/6/2695/18655086/2695\_1\_online.pdf} {https://pubs.aip.org/aip/jap/article-pdf/73/6/2695/18655086/2695\_1\_online.pdf} \BibitemShut {NoStop}%
\bibitem [{\citenamefont {Fiorentino}\ \emph {et~al.}(2007)\citenamefont {Fiorentino}, \citenamefont {Spillane}, \citenamefont {Beausoleil}, \citenamefont {Roberts}, \citenamefont {Battle},\ and\ \citenamefont {Munro}}]{Fiorentino:07}%
  \BibitemOpen
  \bibfield  {author} {\bibinfo {author} {\bibfnamefont {M.}~\bibnamefont {Fiorentino}}, \bibinfo {author} {\bibfnamefont {S.~M.}\ \bibnamefont {Spillane}}, \bibinfo {author} {\bibfnamefont {R.~G.}\ \bibnamefont {Beausoleil}}, \bibinfo {author} {\bibfnamefont {T.~D.}\ \bibnamefont {Roberts}}, \bibinfo {author} {\bibfnamefont {P.}~\bibnamefont {Battle}}, \ and\ \bibinfo {author} {\bibfnamefont {M.~W.}\ \bibnamefont {Munro}},\ }\bibfield  {title} {\enquote {\bibinfo {title} {Spontaneous parametric down-conversion in periodically poled ktp waveguides and bulk crystals},}\ }\href {\doibase 10.1364/OE.15.007479} {\bibfield  {journal} {\bibinfo  {journal} {Opt. Express}\ }\textbf {\bibinfo {volume} {15}},\ \bibinfo {pages} {7479--7488} (\bibinfo {year} {2007})}\BibitemShut {NoStop}%
\bibitem [{\citenamefont {Aspnes}\ and\ \citenamefont {Studna}(1983)}]{PhysRevB.27.985}%
  \BibitemOpen
  \bibfield  {author} {\bibinfo {author} {\bibfnamefont {D.~E.}\ \bibnamefont {Aspnes}}\ and\ \bibinfo {author} {\bibfnamefont {A.~A.}\ \bibnamefont {Studna}},\ }\bibfield  {title} {\enquote {\bibinfo {title} {Dielectric functions and optical parameters of si, ge, gap, gaas, gasb, inp, inas, and insb from 1.5 to 6.0 ev},}\ }\href {\doibase 10.1103/PhysRevB.27.985} {\bibfield  {journal} {\bibinfo  {journal} {Phys. Rev. B}\ }\textbf {\bibinfo {volume} {27}},\ \bibinfo {pages} {985--1009} (\bibinfo {year} {1983})}\BibitemShut {NoStop}%
\bibitem [{\citenamefont {Shoji}\ \emph {et~al.}(1997)\citenamefont {Shoji}, \citenamefont {Kondo}, \citenamefont {Kitamoto}, \citenamefont {Shirane},\ and\ \citenamefont {Ito}}]{Shoji:97}%
  \BibitemOpen
  \bibfield  {author} {\bibinfo {author} {\bibfnamefont {I.}~\bibnamefont {Shoji}}, \bibinfo {author} {\bibfnamefont {T.}~\bibnamefont {Kondo}}, \bibinfo {author} {\bibfnamefont {A.}~\bibnamefont {Kitamoto}}, \bibinfo {author} {\bibfnamefont {M.}~\bibnamefont {Shirane}}, \ and\ \bibinfo {author} {\bibfnamefont {R.}~\bibnamefont {Ito}},\ }\bibfield  {title} {\enquote {\bibinfo {title} {Absolute scale of second-order nonlinear-optical coefficients},}\ }\href {\doibase 10.1364/JOSAB.14.002268} {\bibfield  {journal} {\bibinfo  {journal} {J. Opt. Soc. Am. B}\ }\textbf {\bibinfo {volume} {14}},\ \bibinfo {pages} {2268--2294} (\bibinfo {year} {1997})}\BibitemShut {NoStop}%
\bibitem [{\citenamefont {Peralta~Amores}\ and\ \citenamefont {Swillo}(2022)}]{doi:10.1021/acsanm.1c04202}%
  \BibitemOpen
  \bibfield  {author} {\bibinfo {author} {\bibfnamefont {A.}~\bibnamefont {Peralta~Amores}}\ and\ \bibinfo {author} {\bibfnamefont {M.}~\bibnamefont {Swillo}},\ }\bibfield  {title} {\enquote {\bibinfo {title} {Low-temperature bonding of nanolayered ingap/sio2 waveguides for spontaneous-parametric down conversion},}\ }\href {\doibase 10.1021/acsanm.1c04202} {\bibfield  {journal} {\bibinfo  {journal} {ACS Applied Nano Materials}\ }\textbf {\bibinfo {volume} {5}},\ \bibinfo {pages} {2550--2557} (\bibinfo {year} {2022})},\ \Eprint {http://arxiv.org/abs/https://doi.org/10.1021/acsanm.1c04202} {https://doi.org/10.1021/acsanm.1c04202} \BibitemShut {NoStop}%
\bibitem [{\citenamefont {Peralta~Amores}\ and\ \citenamefont {Swillo}(2024)}]{PhysRevA.110.063713}%
  \BibitemOpen
  \bibfield  {author} {\bibinfo {author} {\bibfnamefont {A.}~\bibnamefont {Peralta~Amores}}\ and\ \bibinfo {author} {\bibfnamefont {M.}~\bibnamefont {Swillo}},\ }\bibfield  {title} {\enquote {\bibinfo {title} {Tunable counterpropagating twin photon source},}\ }\href {\doibase 10.1103/PhysRevA.110.063713} {\bibfield  {journal} {\bibinfo  {journal} {Phys. Rev. A}\ }\textbf {\bibinfo {volume} {110}},\ \bibinfo {pages} {063713} (\bibinfo {year} {2024})}\BibitemShut {NoStop}%
\bibitem [{\citenamefont {Amores}\ and\ \citenamefont {Swillo}(2024)}]{Amores:24}%
  \BibitemOpen
  \bibfield  {author} {\bibinfo {author} {\bibfnamefont {A.~P.}\ \bibnamefont {Amores}}\ and\ \bibinfo {author} {\bibfnamefont {M.}~\bibnamefont {Swillo}},\ }\bibfield  {title} {\enquote {\bibinfo {title} {Heterogeneously integrated ingap/si waveguides for nonlinear photonics},}\ }\href {\doibase 10.1364/OE.520643} {\bibfield  {journal} {\bibinfo  {journal} {Opt. Express}\ }\textbf {\bibinfo {volume} {32}},\ \bibinfo {pages} {16925--16934} (\bibinfo {year} {2024})}\BibitemShut {NoStop}%
\end{thebibliography}%
\end{document}